\documentstyle[preprint,aps,multicol,graphicx]{revtex} 

\draft
\begin{document}

\title{Observation of Thermodynamical Properties in the $^{162}$Dy, $^{166}$Er and $^{172}$Yb Nuclei}

\author{E.~Melby, L.~Bergholt, M.~Guttormsen, M.~Hjorth-Jensen, 
F.~Ingebretsen, S.~Messelt, J.~Rekstad, A.~Schiller, S.~Siem, and S.W.~{\O}deg{\aa}rd}
\address{Department of Physics, University of Oslo, 
Box 1048 Blindern, N-0316 Oslo, Norway}

\maketitle

\begin{abstract}
The density of accessible levels in the ($^3$He,$\alpha \gamma$) reaction 
has been extracted for the $^{162}$Dy, $^{166}$Er and $^{172}$Yb nuclei. The nuclear temperature is measured as a function of excitation energy in the region of 0 -- 6 MeV. The temperature curves reveal structures indicating new degrees of freedom. The heat capacity of the nuclear system is 
discussed within the framework of a canonical ensemble. \end{abstract}


A challenging goal in nuclear physics is to trace thermodynamical quantities as functions of excitation energy. These quantities depend on statistical properties in the nuclear many body system and may reveal phase transitions. Unfortunately, it is difficult to investigate these aspects - both experimentally and theoretically.

The density of levels as a function of excitation energy is the starting point to extract quantities like entropy, temperature and heat capacity. In the pioneer work of Bethe \cite{1} the level density was described within the Fermi gas model using a partition function for the grand-canonical ensemble. This picture of the nucleus as a gas of non-interacting fermions confined to the nuclear volume has later been modified. The phenomenological back-shifted Fermi gas model \cite{2} is a popular extension of the model, which simulates shell and pair correlation effects. This model works well at excitation energies where the level density is high, typically above the neutron binding energy.

In addition to reveal statistical properties of nuclear matter, knowledge of the level density is important in nuclear astrophysics. The level density is essential for the understanding of the nucleosynthesis in stars, where thousands of cross-sections have to be included in the calculations \cite{3}. In parallel with these applications new theoretical approaches are emerging. Recent microscopic model calculations \cite{3} include pair correlations as well as shell effects. With the recent shell model Monte Carlo method \cite{4,5,6,7} one is able to estimate level densities in heavy nuclei up to high excitation energies.

The theoretical progress has so far not been followed by new experimental data. The straightforward way to determine level densities is by counting discrete levels. However, this technique is restricted to light nuclei and/or low excitation energies where the experimental resolution is high enough to resolve individual lines in the spectra. In rare earth nuclei, the estimates obtained by counting levels are only valid up to $\sim$2 MeV of excitation energy. A very useful experimental quantity is the average level spacing observed in slow neutron resonance capture. From these spacings rather accurate level densities for a certain spin value can be determined at the neutron binding energy region. In addition, the level density can be extracted from the shape of continuum particle spectra. However, if pre-equilibrium particle emission takes place, this procedure may be doubtful since high-energy particles reveal high temperatures and, thus, too low level densities.

Recently, the Oslo group has presented a new way of extracting level densities from measured $\gamma$-ray spectra \cite{8,9}. The main advantage of this method is that the nuclear system is very likely thermalized prior to the $\gamma$-ray emission. In addition, the method allows the simultaneous extraction of level density and the $\gamma$-strength function over a wide energy region. In this letter we report for the first time on experimentally deduced temperatures and heat capacities of rare earth nuclei in the 0 -- 6 MeV excitation energy region. 

The experiments were carried out with 45 MeV $^3$He-projectiles at the Oslo Cyclotron Laboratory (OCL). The experimental data are obtained with the CACTUS multidetector array \cite{10} using the ($^3$He,$\alpha \gamma$) reaction on $^{163}$Dy, $^{167}$Er and $^{173}$Yb self-supporting targets. The charged ejectiles were detected with eight particle telescopes placed at an angle of 45$^{\circ}$ relative to the beam direction. An array of 28 NaI $\gamma$-ray detectors with a total efficiency of $\sim$15\% surrounded the target and particle detectors. 

The assumptions behind the method of data analysis and techniques are described in Refs.~\cite{8,9} and only a few comments pertinent to the present work are made here. The experimental level density is deduced from $\gamma$-ray spectra recorded at a number of initial excitation energies $E$, determined by the measured $\alpha$-energy. These data are the basis for making the first-generation (or primary) $\gamma$-ray matrix, which is factorized according to the Brink-Axel hypothesis \cite{11,12} as \begin{equation}
P(E,E_{\gamma}) = \sigma (E_{\gamma}) \rho (E -E_{\gamma}). \end{equation}
From this expression the $\gamma$-ray energy dependent function $\sigma$ as well as the level density $\rho$ is deduced by an iteration procedure [9], using the same 0th-order trial functions as described in Ref.~[9], and the average slopes and the absolute density depend on this starting point. In the following we concentrate only on the level density and its fine structure, which is assumed to be independent of particular $\gamma$-ray decay routes. 

The extracted level densities for the $^{162}$Dy, $^{166}$Er and $^{172}$Yb nuclei are shown as data points in Fig.~1. The data for $^{162}$Dy and $^{172}$Yb deviate slightly from the results previously published~\cite{9}. In the present work the extraction procedure has been refined by omitting data with $E_{\gamma} <$ 1 MeV, where the first-generation $\gamma$-spectra exhibit methodical uncertainties. By excluding these data points, the error bars for the resulting level density are reduced. 

The level density $\rho(E)$ is proportional to the number of states accessible to the nuclear system at excitation energy $E$. Thus, the entropy in the microcanonical ensemble is given by \begin{equation}
S(E)=S_0 + \ln \rho(E),
\end{equation}
where, for convenience, the Boltzmann constant is set to unity ($k_B$ =1). The normalization constant $S_0$ is not important in this discussion, since it vanishes in the evaluation of the temperature \begin{equation}
T(E)=\frac{1}{(\partial S/\partial E)_V}. \end{equation}

Small statistical deviations in the entropy $S$ may give rise to large contributions in the temperature $T$. In order to reduce this sensitivity, the differentiation of $S$ is performed by a least square fit of a straight line to five adjacent data points at a time. The slope of the straight line is taken as the differential of $S$ at that energy. Thus, an effective smoothing of about 0.5 MeV is performed through this procedure. Since the energy particle resolution is around 0.3 MeV, this differentiation procedure will not significantly reduce the potential experimental information. 

The temperatures deduced are shown as discrete data points in Fig.~2. The data reveal several broad structures in the 1 -- 5 MeV region, which are not explained in a Fermi gas description. The most pronounced bumps are located at 1.8 MeV and 3.2 MeV in $^{162}$Dy, at 2.5 MeV and 3.7 MeV in $^{166}$Er and at 1.8 MeV and 3.0 MeV in $^{172}$Yb. These structures are interpreted as being the breaking of nucleon pairs and, at higher energies, the possible quenching of pair correlations. 

The extraction of specific heat capacity is given by \begin{equation}
C_V(E) =\frac{1}{(\partial T/\partial E)_V}, \end{equation}
which effectively means that the entropy $S(E)$ has to be differentiated twice. From the scattered data points of Fig.~2, one immediately sees that this will not give meaningful information for $C_V$ with the present experimental statistics.

At this point one could introduce a strong smoothing of the temperature curve in order to extract $C_V$. However, it seems rather incidental how this should be done. One way to proceed is to introduce the canonical ensemble in the description of the nuclear system. Since the temperature enters as a fixed parameter in this formalism, the statistical uncertainty of $T$ does not introduce any additional fluctuations. The cost of this nice feature is that one has to make an average over a wide excitation energy region.

The partition function in the canonical ensemble \begin{equation}
Z(T)=\sum_{n=0}^{\infty}\rho (E_n)e^{-E_n/T} \end{equation}
is determined by the multiplicity of states at a certain energy. Experimentally, the multiplicity corresponds to the level density of accessible states, $\rho(E_n)$, in the present nuclear reaction at energy bin $E_n$. The widths of the energy bins are 120, 120 and 96 keV for the $^{162}$Dy, $^{166}$Er and $^{172}$Yb nuclei, respectively. 

The mathematical justification of Eq.~(5) is that the sum is performed from zero to infinity. The experimental information on $\rho$ covers the excitation region 0 -- 7 MeV, only. For the three nuclei investigated, the proton and neutron binding energies are $\sim$8 MeV. For excitation energies above this binding energy it is reasonable to assume Fermi gas properties, since single particles may be excited into a region with a very high level density (continuum). Therefore, due to lack of experimental data above 7 MeV, the level density is extrapolated from 7 MeV to higher energies by the Fermi gas model expression \cite{12} \begin{equation}
\rho_{\scriptscriptstyle {FG}}(E)=CE^{-5/4}e^{2 \sqrt{aE}}, \end{equation}
where $C$ is a normalization factor and $a$ is the level density parameter. The best fits of $\rho_{\scriptscriptstyle {FG}}$ to data in the 3.5 -- 6.0 MeV excitation region give a level density parameter of $a$ =17.8, 19.0, and 18.7 MeV$^{-1}$ for the $^{162}$Dy, $^{166}$Er and $^{172}$Y nuclei, respectively. These values are in agreement with data from slow neutron resonances \cite{2} and the semiempirical formula for $a$ with values in between $A/10$ and $A/8$ MeV$^{-1}$. The theoretical level density functions are displayed as solid lines in Fig.~1 and are drawn from 3.5 MeV in order to visualize the fits to data.

In order to determine how far the sum of Eq.~(5) has to be performed, the probability density function
\begin{equation}
p (E)=\frac{\rho_{\scriptscriptstyle {FG}} (E)e^{-E/T}} {\int{\rho_{\scriptscriptstyle {FG}}
(E)e^{-E/T}dE}}
\end{equation}
is shown in Fig.~3 for three typical temperatures. At low temperatures, say $T$ = 0.3 MeV, the nucleons are scattered no higher than $E$ =5 MeV. However, this upper limit increases very rapidly for higher temperatures. For the typical temperatures studied in this letter, $E$ =30 MeV has been tested and found to be a sufficiently high upper limit for the summing in Eq.~(5).

The excitation energy in the canonical
ensemble is given by the thermal average \begin{equation}
<E(T)>=Z^{-1}\sum_{n=0}^{\infty}E_n\rho (E_n)e^{-E_n/T}, \end{equation}
where $\rho$ is the level density (shown in Fig.~1) composed of an experimental and a theoretical part for the excitation regions below and above $\sim$7 MeV, respectively.

The smoothing effect implied by the canonical ensemble can be investigated by calculating the standard deviation $\sigma_{\scriptscriptstyle E}$ for the thermal average of the energy
\begin{equation}
\sigma_{\scriptscriptstyle E}= \sqrt{<E^2>-<E>^2}, \end{equation}
giving e.g. $\sigma_{\scriptscriptstyle E}$ = 3 MeV at $E$ =7 MeV. This again shows that the energy in the canonical ensemble is strongly smoothed for a given temperature. Thus, one cannot expect to find abrupt changes in the thermodynamical quantities. 

From Eq.~(8) the temperature in the canonical ensemble can be studied as a function of $<E>$. In Fig.~2 the canonical ensemble gives a temperature dependence (solid lines) that coincides well with the average values found in the microcanonical ensemble (data points). 

This gratifying behaviour of the canonical temperature encourages us to use the canonical ensemble to estimate the heat capacity $C_V$ as well. The heat capacity can be deduced by simply calculating the increase in the thermal average of the energy $<E>$ with respect to $T$ \begin{equation}
C_V(T)=\frac{\partial <E>}{\partial T}.
\end{equation}

The deduced heat capacities for the $^{162}$Dy, $^{166}$Er and $^{172}$Yb nuclei as functions of $<E(T)>$ are shown in Fig.~4. All nuclei display similar dependencies, reflecting that the thermal averages of the excitation energies smear out structures seen in the experimental level densities. The heat capacities are mainly following the theoretical values obtained by using Eq.~(6) with the proper values of $a$ deduced from the fits of Fig.~1. For comparison, we also show the simplified expression $C_V=2\sqrt{aE}$ assuming $\rho \sim \exp (2 \sqrt{aE})$ with $a =18.5$ MeV$^{-1}$. The canonical heat capacity shows no traces of the fine structures found in the level density below $\sim$4 MeV of excitation energies, as also was the case for the canonical temperature. 

In conclusion, for the first time temperature and heat capacity based on $\gamma$-ray spectra have been extracted for rare earth nuclei. The extracted temperature curves deduced for the microcanonical ensemble reveal structures of unknown origin. The structures in the 1 -- 5 MeV excitation region are interpreted as the breaking of nucleon pairs and quenching of the pair correlations. The heat capacity could only be extracted by the use of the canonical ensemble. These semiexperimental values, which are not expected to show any fine structures, are in agreement with the Fermi gas predictions. It would be very interesting to see if more realistic theoretical calculations can describe the fine structures observed in the microcanonical ensemble.

The authors are grateful to E.A.~Olsen and J.~Wikne for providing the excellent experimental conditions. We wish to acknowledge the support from the Norwegian Research Council (NFR).


\begin{figure}
\includegraphics[totalheight=17.5cm,angle=0,bb=0 80 350 730]{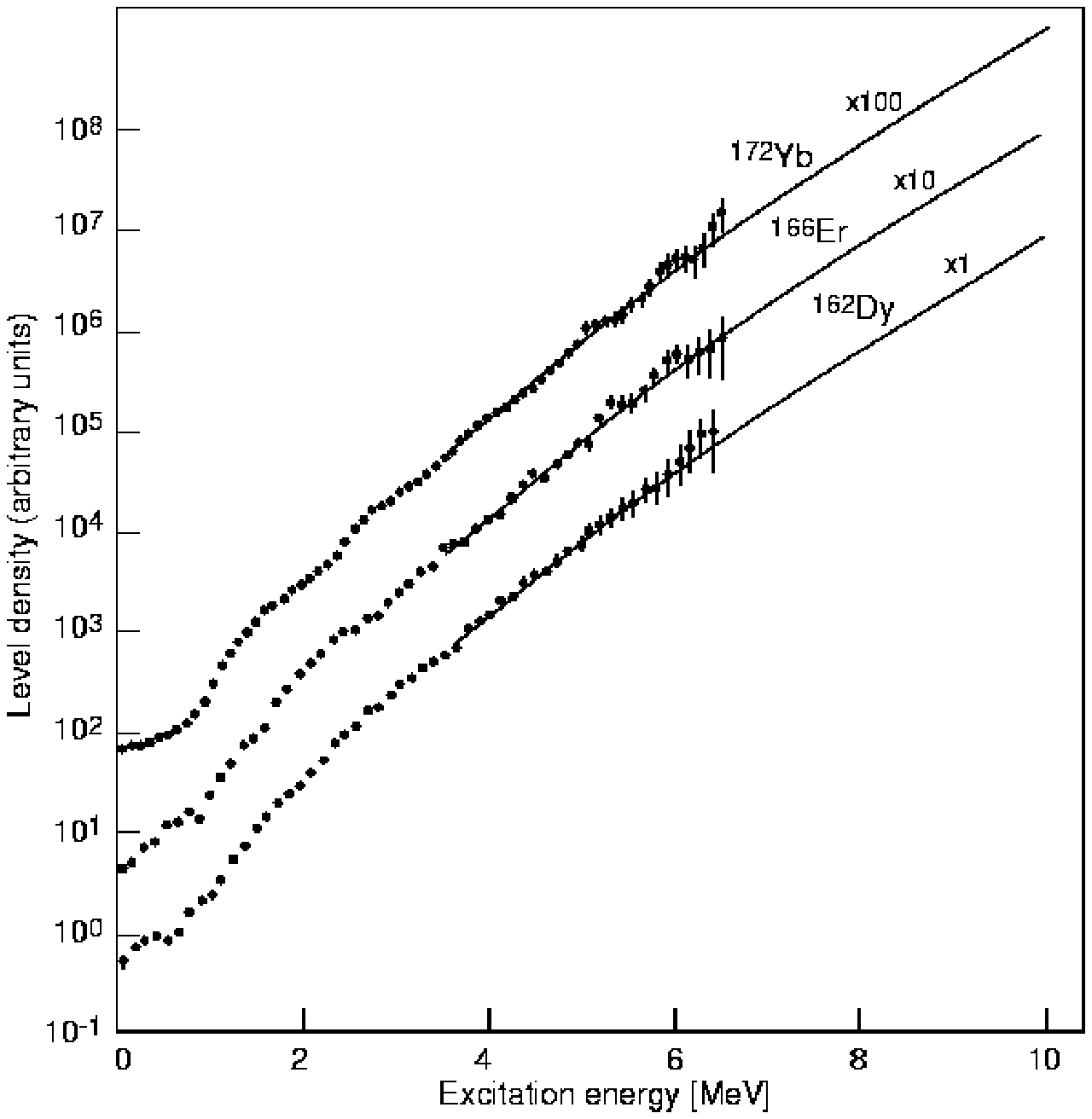} 
\caption{ Extracted level density (points) for $^{162}$Dy, $^{166}$Er and $^{172}$Yb. The error bars show the statistical uncertainties. The solid lines are extrapolations based on the Fermi gas model. The curves are in arbitrary units and separated with a factor of $\sim$10 for better visualization.} \end{figure}

\begin{figure}
\includegraphics[totalheight=19cm,angle=0,bb=0 20 350 730]{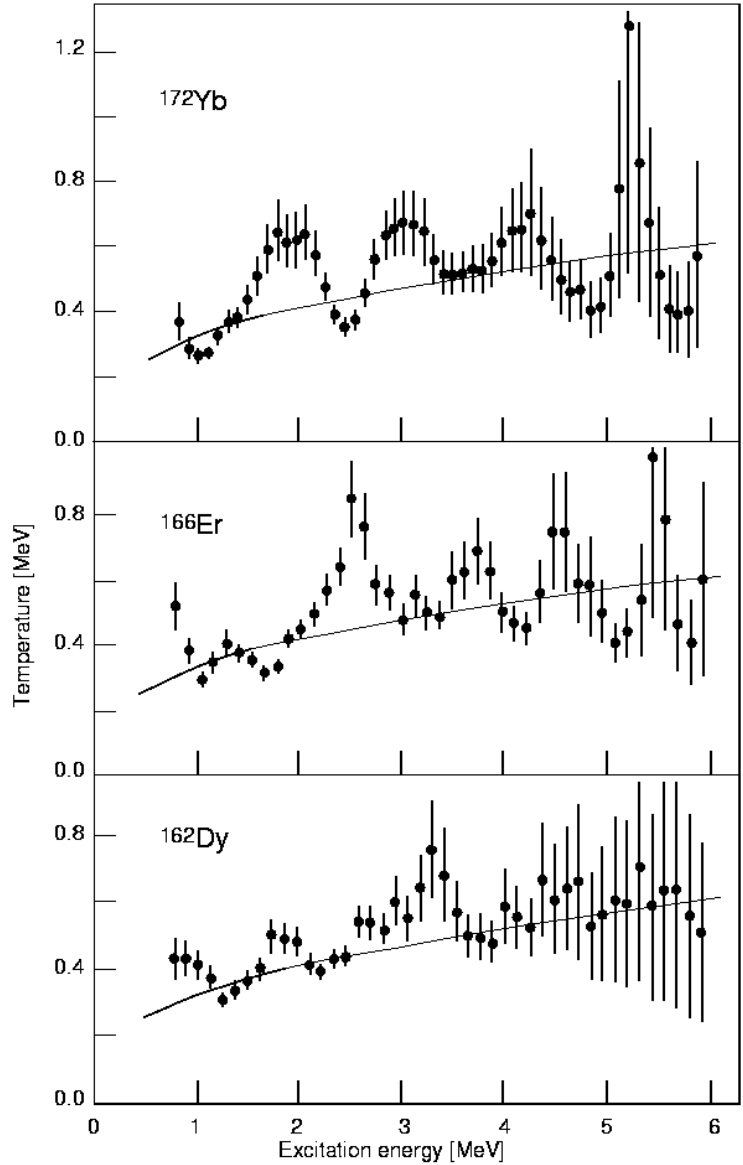} 
\caption{Observed temperatures as functions of the excitation energy $E$ (data points with statistical error bars). The solid lines are temperatures as functions of average excitation energies $<E>$ deduced within the canonical ensemble.} 
\end{figure}

\begin{figure}
\includegraphics[totalheight=17.5cm,angle=0,bb=0 80 350 730]{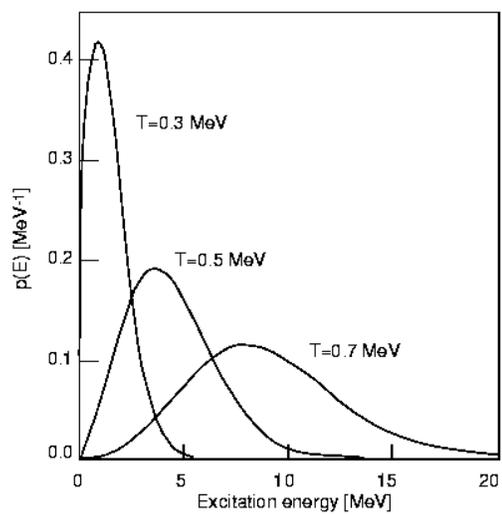} 
\caption{The Fermi gas model energy distribution $p(E)$ of the canonical ensemble.} 
\end{figure}

\begin{figure}
\includegraphics[totalheight=17.5cm,angle=0,bb=0 80 350 730]{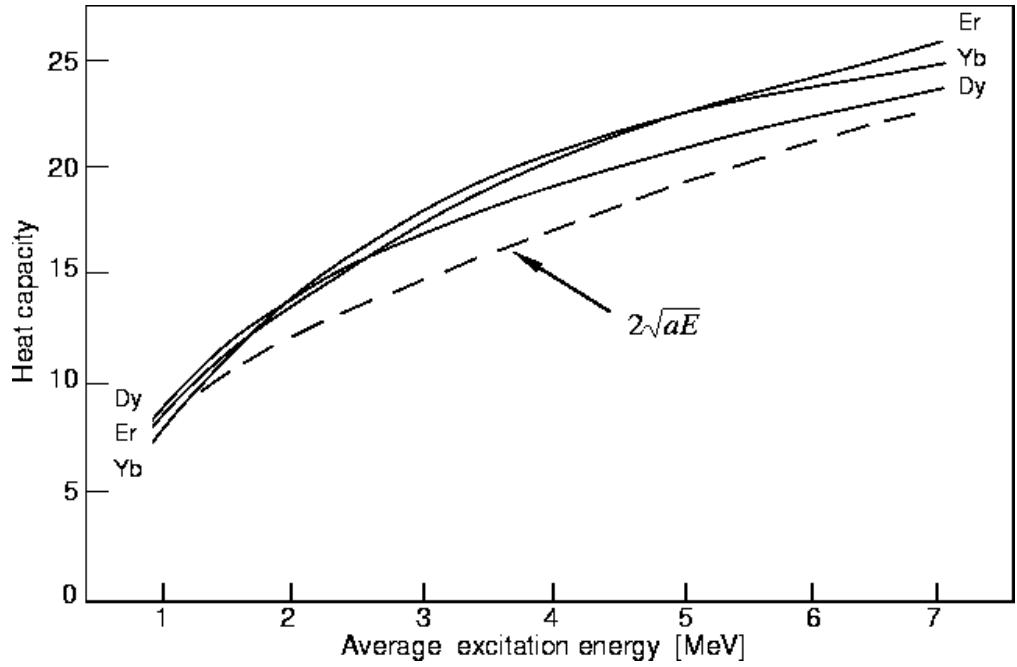} 
\caption{The heat capacities $C_V$ extracted within the canonical ensemble. The dashed curve displays the simplified Fermi gas expression for $a$ =18.5 MeV$^{-1}$.} 
\end{figure}

\end{document}